\documentclass[epj]{svjour}

\usepackage{graphicx}
\usepackage[utf8]{inputenc}
\usepackage[T1]{fontenc}
\usepackage[english]{babel}
\usepackage{braket}
\usepackage{multirow}
\usepackage{dsfont}
\usepackage{mathtools}
\usepackage{xcolor}
\usepackage[normalem]{ulem}

\newcommand{\beq}{\begin{equation}}
\newcommand{\eeq}{\end{equation}}
\newcommand{\bea}{\begin{eqnarray}}
\newcommand{\eea}{\end{eqnarray}}
\newcommand{\nn}{\nonumber}
\newcommand{\eq}{Eq.~}
\newcommand{\eqs}{Eqs.~}
\newcommand{\fig}{Fig.~}
\newcommand{\figs}{Figs.~}
\newcommand{\Sec}{Sec.~}

\newcommand{\bx}{\boldsymbol{x}}

\newcommand{\bp}{\boldsymbol{p}}

\renewcommand{\vec}[1]{\mbox{\boldmath$#1$\unboldmath}}

\def\lsi{\raise0.3ex\hbox{$<$\kern-0.75em\raise-1.1ex\hbox{$\sim$}}}
\def\gsi{\raise0.3ex\hbox{$>$\kern-0.75em\raise-1.1ex\hbox{$\sim$}}}

\begin{document}

\title{Chiral spin symmetry and the QCD phase diagram}

\author{Leonid~Ya.~Glozman\inst{1} \and Owe Philipsen\inst{2}\and Robert D. Pisarski\inst{3}
}                     
\institute{Institute of Physics, University of Graz, A-8010 Graz, Austria
\and
Institute for Theoretical Physics, Goethe University Frankfurt,\\
 Max-von-Laue-Str.\ 1, D-60438 Frankfurt am Main, Germany
 \and 
  Physics Department, Brookhaven National Laboratory, Upton, NY 11973, USA
  }
%
%
\abstract{
Lattice QCD simulations with chirally symmetric quarks have recently established approximate 
$SU(2)_{CS}$ and $SU(2N_F)$ symmetries of the quantum effective action in a temperature range 
above the chiral crossover $T_\mathrm{ch}$, in which color-electric interactions between quarks dominate the dynamics.
We show that such an 
intermediate temperature range between the chirally broken and plasma regimes
is fully consistent with published screening mass spectra, which demonstrate the breakdown of
thermal perturbation theory at the crossover  between the partonic and the chiral spin symmetric regime
at $T_s\sim (2-3)T_\mathrm{ch}$. From the known behavior of screening masses with baryon chemical
potential, we deduce qualitatively how this chiral spin symmetric band extends into the 
QCD phase diagram. In the cold and dense region, we propose parity doubled baryons as possible candidates
for chiral spin symmetric matter. This represents a special case of quarkyonic matter with confinement and restored chiral
symmetry, and can smoothly transform to quark matter at sufficiently high densities.
Finally, we discuss the potential of dilepton spectra to identify such matter forms. 
%
} 
\maketitle

\section{Introduction}	

The theoretical determination of the QCD phase diagram as a function of temperature $T$ and baryon chemical potential
$\mu_B$ remains a challenging task. Despite the persistent sign problem,
combining increasingly detailed lattice information on baryon number 
fluctuations \cite{Borsanyi:2021sxv,Bellwied:2021nrt,Bollweg:2022rps} 
and their modeling \cite{Vovchenko:2017gkg}, the chiral transition in the massless 
limit  \cite{HotQCD:2019xnw,Cuteri:2021ikv} and its relation to the physical point \cite{Halasz:1998qr,Hatta:2002sj},
as well as improved reweighted calculations checked against imaginary chemical potential \cite{Borsanyi:2021hbk}, 
one may bound the critical endpoint of a possible chiral
phase transition at finite density by $\mu_B\gsi 2.5T$ and $T\lsi 135$ MeV. 
Critical endpoint candidates predicted by functional methods 
are consistent with and well beyond these bounds in a region, 
where the current truncations require further stability checks \cite{Fischer:2018sdj,Fu:2019hdw,Gao:2020fbl}. For recent reviews and
further references, see \cite{Philipsen:2021qji,Guenther:2022wcr,Fischer:2018sdj}.  
\begin{figure}[t]
    \centering
    \includegraphics[width=0.98\columnwidth, clip, trim=3mm 2mm 0 0]{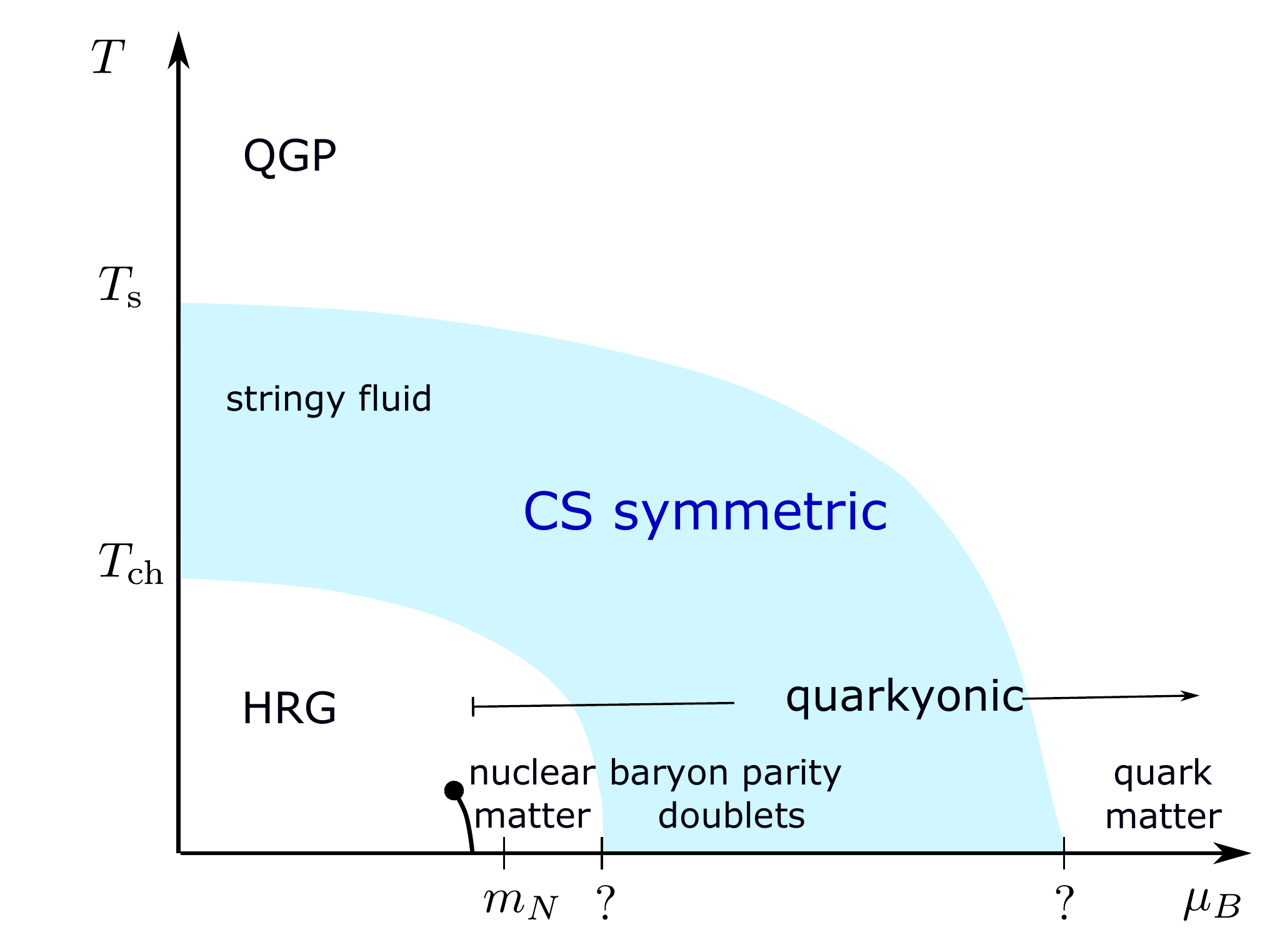}
    \caption{Qualitative sketch of a possible QCD phase diagram with a band of approximate chiral spin symmetry. At zero density, both 
    transitions between the regimes are smooth crossovers. The lower boundary corresponds to full chiral symmetry restoration,
    which could be a true phase transition at larger density. Beyond the upper boundary, light quark meson correlators are consistent
    with a partonic description.}
    \label{fig:PD}
\end{figure}

On the other hand, an unexpected new symmetry was discovered to emerge at zero density, where there is no sign problem.
Lattice simulations with $N_F=2$ flavors of chiral fermions show approximate 
$SU(2)_{CS}$ chiral spin and $SU(2N_F)$ symmetries for both spatial and temporal $J=0,1$ meson correlators in a 
temperature range $T_\mathrm{ch}\lsi T \lsi 3T_\mathrm{ch}$ above the 
chiral crossover, $T_\mathrm{ch}$  \cite{Rohrhofer:2019qwq,Rohrhofer:2019qal}. 
This symmetry is larger than the $SU(2)_L\times SU(2)_R\times U(1)_A$
chiral symmetry of the QCD Lagrangian, which it contains as a subgroup. It is a symmetry of the color charge, but not
of the QCD Lagrangian, and can only be realized  in the quantum effective action 
when color-electric quark gluon interactions dynamically dominate over
color-magnetic interactions and kinetic terms. This suggests that 
chiral quarks are bound by color-electric flux tubes in such a regime, which thus has been dubbed a 
``stringy liquid'' \cite{Rohrhofer:2019qwq,Rohrhofer:2019qal}. When temperature increases beyond
$T_s\sim 3T_\mathrm{ch}$, the color-electric interactions between quarks get screened and the
 symmetry reduces to the expected chiral symmetry corresponding to a quark gluon plasma.

In this work, we elaborate on a suggestion \cite{Philipsen:2019rjq} 
that the chiral spin symmetric temperature range observed at zero density 
continues as a band across the QCD phase diagram,  
sketched in \fig\ref{fig:PD}. To render this paper self-contained, we begin with a brief summary 
of chiral spin symmetry in \Sec\ref{sec:CS} and its observation in finite temperature lattice QCD, \Sec\ref{sec:obs}.
In \Sec\ref{sec:ms} we show that all known results on screening masses are fully consistent with 
such an intermediate temperature range between broken chiral symmetry and a partonic quark gluon plasma.
Using quark hadron duality of screening masses to identify the onset of the plasma regime, we derive how the upper boundary of the
chiral spin symmetric band curves away from the $T$-axis in \Sec\ref{sec:tmu}. In \Sec\ref{sec:pdub} we
identify parity doubled baryon matter as a candidate for a chiral spin symmetric 
regime of cold and dense QCD, which can be naturally embedded into quarkyonic matter.
Finally, we discuss the prospects and limitations of dilepton spectra to probe
matter in the chiral spin symmetric regime, \Sec\ref{sec:dilep}.  

\section{Chiral spin symmetry of the color charge and its implications \label{sec:CS}}
 
 The Banks-Casher relation \cite{Banks:1979yr} connects the quark
 condensate of the QCD vacuum with the density of the near-zero modes $\lambda_n$ of the Dirac operator,
 \bea
 \langle \bar{\psi}\psi\rangle&=&\pi \rho(0)\;,\\
\rho(0) &=&
 \lim_{\lambda \rightarrow 0} \lim_{m \rightarrow 0} \lim_{V\rightarrow \infty} 
\frac{1}{\pi V} \int
d \lambda \sum_n \frac{\delta(\lambda-\lambda_n)}{m + i\lambda}\;.\nn
 \eea
 An artificial truncation of the near-zero modes on the lattice at $T=0$ may then be expected to restore
 the  $SU(N_F)_L \times SU(N_F)_R$ and possibly the $U(1)_A$
 chiral symmetry of the QCD Lagrangian. For example, the instanton liquid model \cite{Shuryak:1981ff,Diakonov:1985eg} 
 suggests that both $SU(N_F)_L \times SU(N_F)_R$ and $U(1)_A$ breakings
 are due to the 't Hooft determinant induced by the instanton fluctuations of the QCD vacuum 
 at sufficiently strong coupling~\cite{tHooft:1986ooh}.

 A spectrum calculation based on such truncated Dirac operators has revealed a larger
 degeneracy pattern than expected, both for mesons \cite{Denissenya:2014poa,Denissenya:2014ywa,Denissenya:2015mqa} 
 and baryons \cite{Denissenya:2015woa}.
 From the quantum numbers of the degenerate states the symmetry groups 
 responsible for this large degeneracy,  the  chiral spin $SU(2)_{CS}$ and $SU(2N_F)$, 
 were reconstructed in refs. \cite{Glozman:2014mka,Glozman:2015qva}.
 An  $SU(2)_{CS}$  chiral spin transformation acting on Dirac spinors 
 can be defined as
 \begin{equation}
\label{eq:V-defsp}
  \psi \rightarrow  \psi^\prime = \exp \left(i  \frac{\varepsilon^n \Sigma^n}{2}\right) \psi \;,
\end{equation}
where the generators $\Sigma^n/2$ of the four-dimensional reducible
representation are
\begin{equation}
 \Sigma^n = \{\gamma_0,-i \gamma_5\gamma_0,\gamma_5\}
\label{SIGCS}
\end{equation}
and satisfy the $su(2)$ algebra. This transformation rotates
in the space of right- and left-handed Weyl spinors $R,L$, and an equivalent representation 
of \eq(\ref{eq:V-defsp}) is
\begin{equation}
\left(\begin{array}{c}
R\\
L
\end{array}\right)\; \rightarrow
\left(\begin{array}{c}
R'\\
L'
\end{array}\right)=
\exp \left(i  \frac{\varepsilon^n \sigma^n}{2}\right) \left(\begin{array}{c}
R\\
L
\end{array}\right)\; .
\label{eq:RL}
\end{equation}

In Euclidean spacetime with its $O(4)$ symmetry, all four directions are 
equivalent and one can use any Euclidean  hermitian $\gamma$-matrix $\gamma_k$, $k=1,2,3,4$ to replace
the Minkowskian $\gamma_0$,
\begin{equation}
 \Sigma^n = \{\gamma_k,-i \gamma_5\gamma_k,\gamma_5\},
\end{equation} 
\begin{equation}
\gamma_i\gamma_j + \gamma_j \gamma_i =
2\delta^{ij}; \qquad \gamma_5 = \gamma_1\gamma_2\gamma_3\gamma_4.
\label{gamma}
\end{equation}
The $su(2)$ algebra 
is satisfied for any $k=1,2,3,4$, so any choice is permitted 
that does not mix operators with different spatial 
$O(3)$ spins.
Note that $SU(2)_{CS}$  contains 
$U(1)_A$ as a subgroup.
The direct product of the $SU(2)_{CS}$ group with the flavor group $SU(N_F)$
can be embedded into a $SU(2N_F)$ group,
which includes the chiral
symmetry as a subgroup,
\begin{equation}
SU(2N_F)\supset SU(N_F)_L \times SU(N_F)_R \times U(1)_A \;.
\label{eq:sym}
\end{equation}
The multiplets of the  $SU(2)_{CS}$ and $SU(4)$ groups
have been worked out in Refs.~\cite{Glozman:2014mka,Glozman:2015qva}. In particular,
these symmetries require the degeneracy of \textit{all} isovector $J=1$
mesons, including the $a_1$ and $b_1$, which are not degenerate under chiral symmetry.

The $SU(2)_{CS}$ and $SU(2N_F)$ groups are not
symmetries of the 
massless Dirac part of the QCD Lagrangian.
In a fixed Lorentz frame we can split the latter in
color-electric (temporal) and color-magnetic (spatial) parts,
\begin{equation}
\bar{\psi}\gamma^\mu D_\mu \psi= \bar{\psi}\gamma^0 D_0 \psi + \bar{\psi}\gamma^i D_i\psi\;,
\end{equation}
where the first term is invariant under $SU(2)_{CS}$ and $SU(2N_F)$, while the second term is not.
At the same time these are symmetries
of the Lorentz-invariant color charge
\begin{equation}
Q^a =  \int d^3x  \;\;
\psi^\dagger(x) T^a  \psi(x)\;,
\label{Q}
\end{equation}
with $T^a$ the $SU(3)$ color generators.
This   feature allows for the 
$SU(2)_{CS}$ and $SU(2N_F)$ symmetries to distinguish between the
chromoelectric and chromomagnetic interactions in a given reference frame.
The chromoelectric gauge field couples to 
the color charge, consequently the chromoelectric interaction 
of quarks and gauge fields is $SU(2)_{CS}$ and $SU(2N_F)$ symmetric. The chromomagnetic gauge fields couple to
a current, which is not 
$SU(2)_{CS}$ and $SU(2N_F)$ symmetric. Thus, the symmetry of the electric part of the
QCD Lagrangian is larger than the symmetry of the QCD Lagrangian as a whole.

The observation of the $SU(2)_{CS}, SU(2N_F)$ symmetries in the hadron spectrum upon
truncation of the near-zero modes of the Dirac operator then implies
that the magnetic interaction at zero temperature is located mostly in those near-zero modes,
whereas a confining electric interaction is distributed among all Dirac modes. 
Hence, confinement and chiral symmetry breaking in QCD are not directly related phenomena. Based on this observation
it was predicted that, for finite temperature QCD without any truncations, 
where the rest frame of the medium constitutes a preferred reference frame,
the chiral spin and $SU(2N_F)$ symmetries should emerge 
above the chiral symmetry restoring crossover \cite{Glozman:2016swy}.
 
\section{Chiral spin symmetry at finite temperature \label{sec:obs}}

\begin{figure*}[t]
    \centering
    \includegraphics[width=0.4\textwidth]{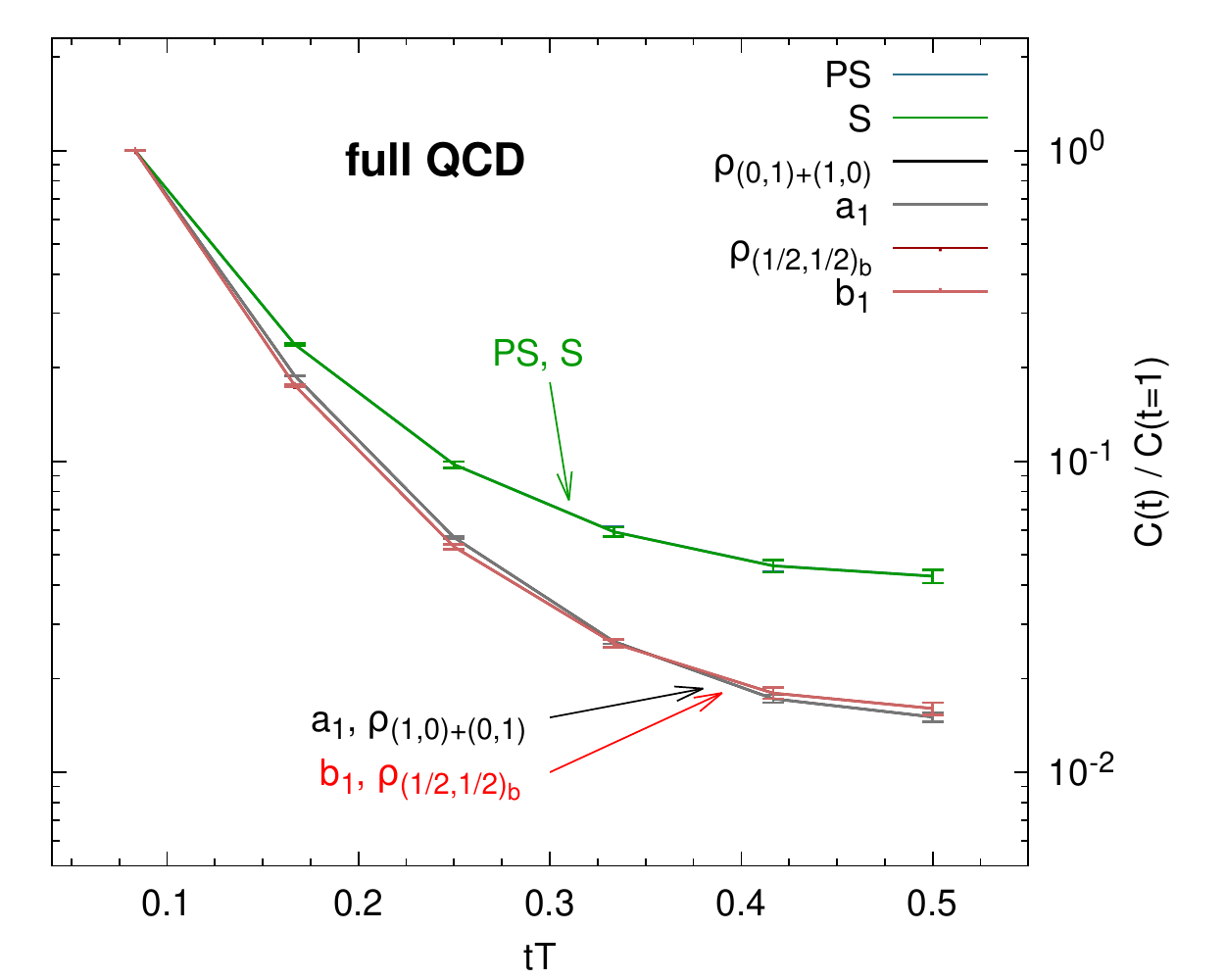}\hspace*{1cm}
    \includegraphics[width=0.4\textwidth]{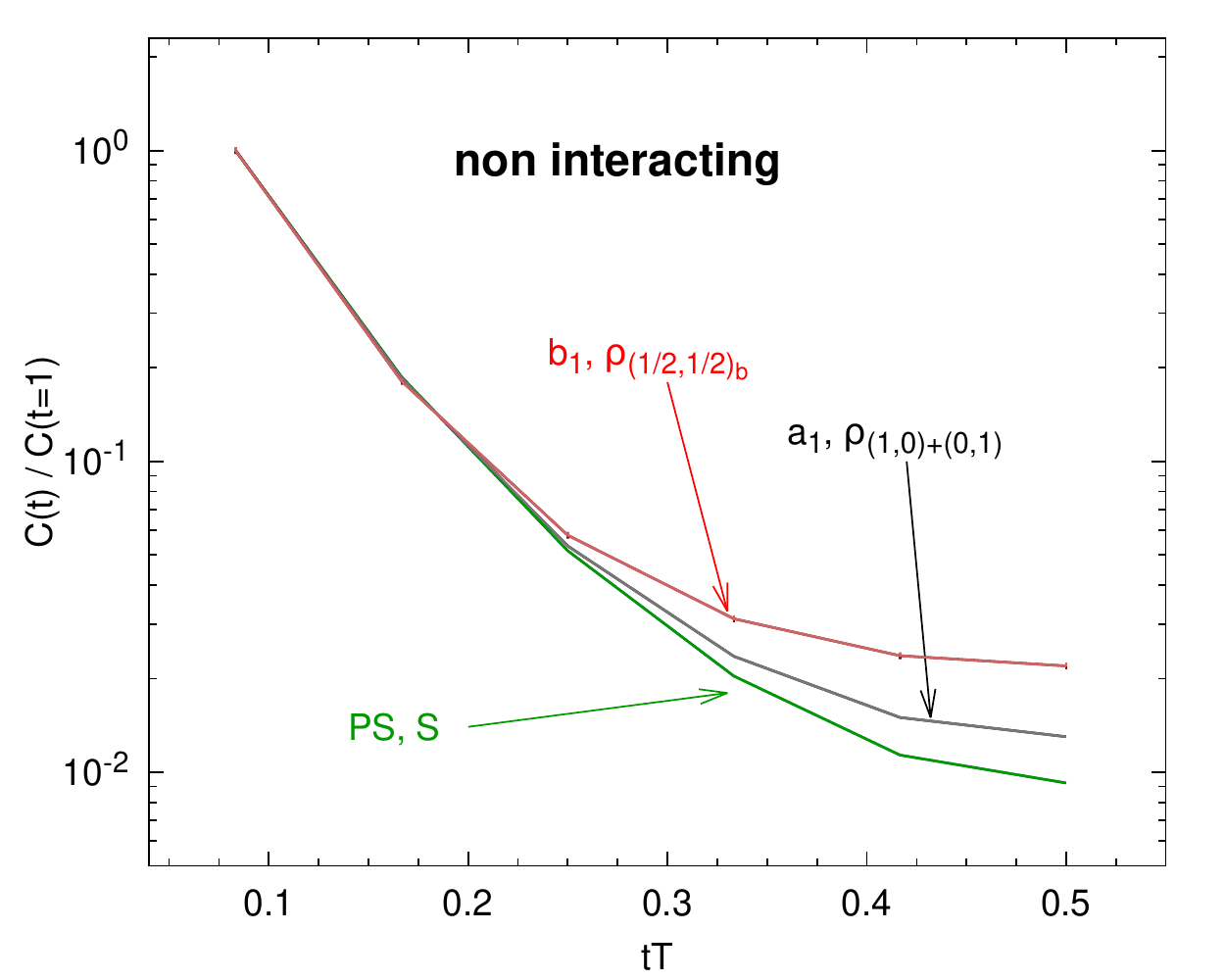}
    \caption[]{
    Temporal correlation functions for $N_F=2$ QCD with chiral fermions on $12 \times 48^3$ lattices. 
    Left: Full QCD results at $T=220$ MeV, representing multiplets of all groups, $U(1)_A$, 
    $SU(2)_L \times SU(2)_R$, $SU(2)_{CS}$ and $SU(4)$.
    Right: Correlators calculated with free quarks with manifest $U(1)_A$ and $SU(2)_L \times SU(2)_R$ symmetries.
   From \cite{Rohrhofer:2019qal}.}
    \label{fig:temp}
\end{figure*}
At finite temperature, the chiral condensate decreases significantly
through a  smooth crossover between $T\approx 100 - 200$ MeV. 
The pseudo-critical temperature for chiral symmetry restoration is usually
defined by the peak of the chiral susceptibility,  
and for $N_F=2+1$ QCD at the physical point in the continuum 
is $T_\mathrm{pc} =157(2)$ MeV \cite{HotQCD:2018pds,Borsanyi:2020fev}.
Above this temperature, one a priori expects observables to exhibit a  
$SU(2)_L \times SU(2)_R$ chiral symmetry.  
The effects of the axial anomaly 
are determined by the topological charge
density. There are strong indications from the lattice that the
$U(1)_A$ symmetry is  approximately restored above 
$T_\mathrm{ch} \approx 200$ MeV \cite{Cossu:2013uua,Tomiya:2016jwr,Bazavov:2019www,Aoki:2020noz},
which suggests that the topological fluctuations at these temperatures
are strongly suppressed. This effective symmetry restoration
is visible by the degeneracy of all correlators (obtained with a
chirally symmetric Dirac operator) connected by the $U(1)_A$ transformation
\cite{Rohrhofer:2019qwq,Rohrhofer:2019qal}, 
and the degeneracy of scalar and pseudo-scalar susceptibilities in particular. 
Closer to $T_\mathrm{pc}$, the quark condensate becomes appreciable
and should provide a splitting of the respective correlators, as is also observed \cite{Bazavov:2019www,Kaczmarek:2021ser}.
For the following, mostly qualitative, considerations, we take $T_\mathrm{ch}\approx T_\mathrm{pc}$ approximately,
without loss of generality.
  
Detailed lattice studies of spatial \cite{Rohrhofer:2019qwq} and temporal \cite{Rohrhofer:2019qal}
meson correlators at $T\gsi T_\mathrm{ch}$, 
calculated in $N_F=2$ QCD with a chirally symmetric Dirac operator
at physical quark masses, exhibit approximate multiplets of both 
$SU(2)_{CS}$ and $SU(2N_F)$ groups, i.e.~they display a symmetry larger than 
the chiral symmetry of the QCD Lagrangian. 
As an example and for later reference, we reproduce the temporal correlators 
from \cite{Rohrhofer:2019qal} in  \fig\ref{fig:temp}.  
Correlators of the isovector scalar ($S$) and isovector pseudo-scalar ($PS$) operators are connected by the $U(1)_A$ 
transformation and their degeneracy indicates an effective restoration of this symmetry. 
If there is a tiny splitting of the $S$ and $PS$ correlators, it is too small to be seen in the present lattice data. 
As per \eq(\ref{eq:sym}), this symmetry is necessary, but not sufficient to demonstrate realisation of the larger chiral spin symmetry.
An approximate degeneracy of the $a_1$, $b_1$, $\rho(1,0)+(0,1)$ and $\rho(1/2,1/2)_b$ correlators indicates emergent approximate
$SU(2)_{CS}$ and $SU(4)$ symmetries.
This larger symmetry
disappears again once temperatures exceed $T\gsi 3 T_\mathrm{ch}$ \cite{Rohrhofer:2019qwq,Rohrhofer:2019qal}.
Let us assess the implications of this observation in some detail.

For any meson operator $O_\Gamma(\tau,\bx)=\bar{\psi}(\tau,\bx)\Gamma \frac{\boldsymbol{\tau}}{2}\psi(\tau,\bx)$ with 
$\Gamma\in\{1,\gamma_5,\gamma_\mu,\gamma_5\gamma_\mu,\sigma_{\mu\nu},\gamma_5\sigma_{\mu\nu}\}$, the Euclidean correlation functions,
\begin{equation}
C_\Gamma(\tau,\bx)=\langle O_\Gamma(\tau,\bx)\,O_\Gamma^\dag(0,\mathbf{0})\rangle\;,
\end{equation}
carry the full spectral information of all excitations with $J=0,1$ in their
associated spectral functions $\rho_\Gamma(\omega,\bp)$,
\begin{eqnarray}
C_\Gamma(\tau,\bp) &=&\int_0^\infty \frac{d\omega}{2\pi}\;K(\tau,\omega)\rho_\Gamma(\omega,\bp)\;, \nonumber \\
K(\tau,\omega)&=&\frac{\cosh(\omega(\tau-1/2T))}{\sinh(\omega/2T)}\;.
\label{eq:corr}
\end{eqnarray}
The spatial and temporal correlators probed in  \cite{Rohrhofer:2019qwq,Rohrhofer:2019qal}, 
\begin{eqnarray}
C_\Gamma^s(z)=\sum_{x,y,\tau} C_\Gamma(\tau,\bx)\;,\label{eq:c_z}\\
C_\Gamma^\tau(\tau)=\sum_{x,y,z}C_\Gamma(\tau,\bx)\;,
\end{eqnarray}
collect the spectral information projected on the $(p_x=p_y=\omega=0)$ and $(p_x=p_y=p_z=0)$  axes, respectively.
In thermal equilibrium the system is isotropic and the momentum distribution is the same in all 
directions, $\rho_\Gamma(\omega,\bp)=\rho_\Gamma(\omega,|\bp|)$. Observing approximate chiral spin symmetry  
both in the frequency and one momentum direction is therefore sufficient to conclude that it is
also approximately realized in the full spectral functions $\rho_\Gamma(\omega,\bp)$. Finally, since 
different quantum number channels are evaluated with the same action,
one must conclude that the observed degeneracy patterns reflect an approximate symmetry
of the non-perturbative effective action, and hence the thermal partition function of QCD.

Finite temperature chiral spin symmetry is thus an example of an emergent symmetry. 
Similar to the synthetic vacuum situation described in the last section, for this to happen the 
chromoelectric sector of the effective quark action must dominate over the chromomagnetic sector.
Moreover, the chromoelectric interaction has to dominate over the spatial kinetic terms, 
which implies that the effective action is far from that of a weakly interacting system.
Indeed, meson correlators evaluated in a free quark gas are even qualitatively incompatible with 
the observed multiplet structure \cite{Rohrhofer:2019qwq,Rohrhofer:2019qal}, as \fig\ref{fig:temp} demonstrates.
This suggests that the degrees of freedom of QCD in the chiral spin symmetric regime, $T_\mathrm{ch}\lsi T \lsi 3 T_\mathrm{ch}$,
are chirally symmetric quarks bound to color singlet objects by the
chromoelectric field. 

Disappearance of these
symmetries for $T\gsi 3 T_\mathrm{ch}$ indicates that the chromoelectric
interactions between light quarks get screened, and one observes a smooth crossover to
a quark gluon plasma with quasiquarks and quasigluons being effective degrees of freedom.
The latter picture is supported by the success 
of the hard thermal loop approach \cite{Haque:2014rua}
at these temperatures.

At zero density, there are then three temperature regimes in QCD with clearly distinguishable symmetries:
the low temperature regime with spontaneously broken chiral symmetry, an intermediate regime with approximate
chiral spin and $SU(2N_F)$ symmetries, and a high temperature regime 
with chiral symmetry\footnote{
There are several other observations of non-perturbative dyna\-mics above $T_\mathrm{ch}$. 
The concept of a semi-QGP \cite{Dumitru:2012fw} predicts a separation of chiral symmetry restoration
and deconfinement by an intermediate $T\sim 155-350$ MeV range \cite{Pisarski:2016ixt}. In recent lattice simulations
at the physical point, thermal monopole
condensation, often interpreted as marking the transition between confined and deconfined regimes, is
observed at $T\approx 275$ MeV \cite{Cardinali:2021mfh}, and the spectral density of a chiral Dirac operator
suggests a novel phase $T\sim 200-250$ MeV with approximate IR scale invariance
\cite{Alexandru:2019gdm,Alexandru:2021pap}. At present it is not clear if and how these phenomena
are related to chiral spin symmetry.}.

\section{Screening masses \label{sec:ms}}

Ultimately, the nature of the degrees of freedom composing the thermal system in its different regimes 
is encoded in the spectral functions. At present, these are not yet available fully non-perturbatively. 
However, we have 
increasingly detailed, non-perturbative knowledge of screening masses, which govern 
the exponential decay of spatial correlators, \eq(\ref{eq:c_z}). For the following it is useful to recall that,
on a Euclidean space time lattice, the thermal partition function can be represented equivalently by two different
Hamiltonians,
\begin{eqnarray}
e^{pV/T}=Z&=&{\rm Tr}(e^{-aHN_\tau})=\sum_{n}e^{-aE_nN_\tau} \nonumber\\
&=&{\rm Tr}(e^{-aH_zN_z})=\sum_{n_z}e^{-aE_{n_z}N_z}\;.
\label{eq:ham}
\end{eqnarray} 
Here,
$H$ is the usual QCD Hamiltonian translating states by one
lattice spacing in Euclidean time,  whereas $H_z$ is the analogous operator translating states in the $z$-direction,
\bea
|\psi(\tau+1;\bx)\rangle &=&\exp(-aH)|\psi(\tau;\bx)\rangle\;,\nn\\
|\psi(\tau; x,y,z+1)\rangle &=&\exp(-aH_z)|\psi(\tau;\bx)\rangle\;.
\eea
On the lattice, both are straightforwardly defined without gauge fixing via the lattice action between adjacent
$\tau$- or $z$-slices, respectively \cite{Montvay:1994cy}.

The thermodynamic limit ($N_{x,y,z}\rightarrow \infty$ with $T^{-1}=aN_\tau$ finite) 
formally represents the ``vacuum'' physics of $H_z$, whose 
spectrum is sensitive to the ''finite volume effect'' of the compactified $\tau$-direction, i.e.~$T^{-1}$. 
The screening masses are the corresponding ground state energies 
in each quantum number channel. Obviously, in the limit $T=0$ the spectrum is identical to that of $H$, while for 
$T\rightarrow\infty$ it reduces to the spectrum of 
3d QCD, which is known as dimensional reduction. 
Evidently, screening masses are directly related to the equation of state, which is completely
determined by the full spectrum of $H_z$.

In order to characterize the dominant dynamical degrees of freedom, it is natural to proceed in analogy to
vacuum QCD, where rarely any confusion arises between hadronic physics and quark gluon physics. 
While experimental initial and final states are ever exclusively hadronic, one may speak of parton physics driving the dynamics
whenever quark hadron duality holds \cite{Shifman:2000jv}, i.e.~the hadronic observables follow perturbative predictions for
partonic (sub-~)~processes. This is also the terminology adopted in some discussions of experimental results, 
see e.g.~\cite{Andronic:2017pug}. For a thermal equilibrium system, 
screening masses are accessible by perturbative and non-perturbative calculations, 
thus providing a viable theoretical testing ground. 

\subsection{Chromoelectric vs. chromomagnetic fields}

Thermal QCD generates three parametrically
distinguished scales, the hard scale of the non-zero Matsubara modes, $\sim \pi T$, the intermediate scale of
the color-electric fields, $\sim gT$, and the fully non-perturbative soft scale $\sim g^2T$ of the color-magnetic 
fields \cite{Kapusta:2006pm,Bellac:2011kqa}. For sufficiently small gauge coupling, the scales 
are separated and the harder modes can be integrated out to successively produce the effective theories EQCD, describing   
the gauge fields $A_0,A_i$ on scales $\lsi gT$, and MQCD for 
$A_i$ on scales $\lsi g^2T$. The latter is equivalent to three-dimensional Yang-Mills theory and fully non-perturbative. 

The balance between color-electric and color-magnetic fields 
was studied on the lattice by a mixing analysis of correlation matrices of gauge invariant gluonic operators 
within EQCD \cite{Hart:2000ha}. 
 At $T\approx 2T_\mathrm{ch}$ the lowest screening mass is 
associated with the operator ${\rm Tr}(A_0^2)$, whereas the one pertaining to ${\rm Tr} (F_{ij}^2)$
is more than twice as large. Hence, the dynamical ordering of ``soft'' and ``ultra-soft'' scales is opposite
to the perturbative expectation. The color-electric fields cannot be integrated out, but 
rather give the largest contributions to the EQCD partition function at this temperature. 
This demonstrates their dynamical dominance in this regime,
and fully supports the emergence of chiral spin symmetry as a consequence of non-perturbative gauge field dynamics.

\subsection{The Debye mass}

According to a non-perturbative definition of the Debye mass based on Euclidean time reflection of gauge 
invariant operators~\cite{Arnold:1995bh}, 
lattice evaluations at $T\approx 2T_\mathrm{ch}$ 
give $m_D^\mathrm{gi}\approx 7.5 T$ \cite{Hart:2000ha,Borsanyi:2015yka,Andreoli:2017zie}, 
which amounts to a Debye radius of $r_D\approx 0.09$~fm.  
Defining the Debye mass instead as the matching coefficient of the $A_0^2$-term in EQCD, which to leading order 
corresponds to the propagator pole mass, 
one obtains $m_D^\mathrm{pole}\approx 2.5 T$  \cite{Laine:2019uua} or $r_D\approx 0.27$~fm. While rather different, both 
definitions result
in a screening length smaller than a typical hadron size.
A chiral spin symmetric regime composed of hadron-like objects 
thus appears to contradict the 
common picture of Debye screening~\cite{Matsui:1986dk}, as was also pointed out in~\cite{Shuryak:2019fgq}.

However, both definitions of the Debye mass are based on pure gauge quantities and related to the screening of static charges.
Even for heavy quarks the dynamical picture is more complicated, with mass values differing widely 
between quantum number channels, and the precise connection between the Debye mass and 
the dissociation of bound states remains far from clear, for a review see \cite{Bazavov:2020teh}. 
In the context of chiral spin symmetry 
we are interested in the fate of the light quarks and mesons, which also give the dominant contribution to the 
equation of state.
But relativistic quarks have no associated
potentials in the first place, and chromoelectric flux distributions within light mesons will depend on all quantum numbers and 
behave quite differently from those between
static quarks.

 Moreover, restricting QCD to $N_f=2+1$ light flavors, as is done in most lattice simulations at the physical point, 
neither propagator poles nor heavy quarkonium 
screening masses enter the partition function \eq(\ref{eq:ham}) at all. 
Only $m^\mathrm{gi}_D$  can possibly appear as screening mass pertaining to
the purely gluonic $J^{PC}=0^{-+}$ operator $\mathrm{Tr} (F_{ij}A_0)$ \cite{Hart:2000ha}. 
This represents one single term, which is subdominant since
its screening mass value is larger than those of all twelve flavor non-singlet $J=0,1$ mesons
to be discussed below. The Debye mass therefore has little influence on the thermodynamics of light quarks. 

\subsection{Meson screening masses}

\begin{figure*}[t]
    \centering
        \includegraphics[width=0.8\columnwidth, clip, trim=3mm 2mm 0 0]{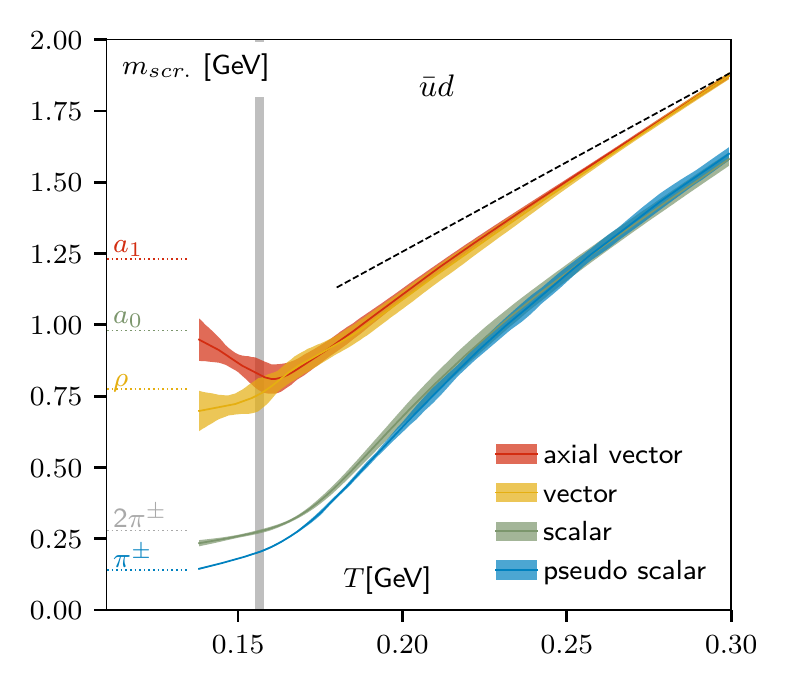}\hspace*{1cm}
    \includegraphics[width=0.8\columnwidth, clip, trim=3mm 2mm 0 0]{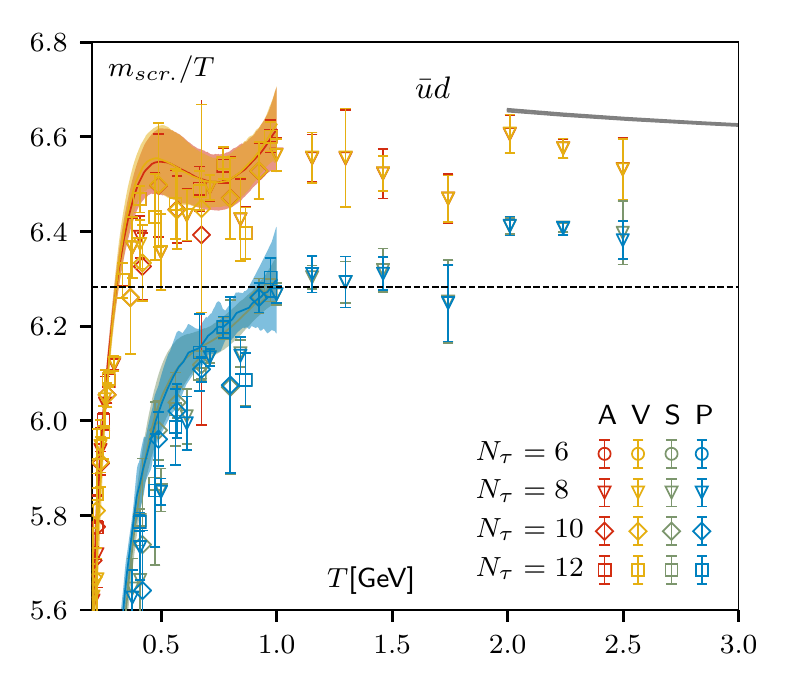}
    \caption[]{Screening masses of the lightest $\bar{u}d$ mesons, evaluated in simulations using 
    HISQ fermions, from \cite{Bazavov:2019www}.}
    \label{fig:screen}
\end{figure*}

What we need to do instead is to study meson screening masses in the light quark sector. A lot of progress has been 
made, both analytically and numerically, towards an increasingly precise evaluation over a wide temperature range. 
In Fig.~\ref{fig:screen}
we reprint a recent lattice determination of the scalar and vector screening masses composed of $\bar{u}d$ 
quarks  \cite{Bazavov:2019www}.
Also shown is the leading perturbative result $\sim 2\pi T$, corresponding to the Matsubara modes of two free quarks, 
and the first correction $\sim g^2$ evaluated within EQCD \cite{Laine:2003bd}. Note that this includes an all-loop-order
HTL resummation of soft contributions from the scale $\sim gT$.  One observes the screening masses in both vector and 
scalar channels to overshoot the $\sim 2\pi T$ level and to slowly approach the $O(g^2)$ prediction, while 
spin dependence enters the perturbative series at $O(g^4)$ only \cite{Koch:1992nx,Hansson:1991kb}.

Lattice calculations of pseudo-scalar and vector meson screening masses have recently been extended with 
unprecedented precision to the high temperature range $T=1-160$~GeV \cite{Brida:2021ytm}, permitting a detailed analysis of 
their perturbative behavior. In particular, over all three orders of magnitude in temperature, the 
lattice data are 
well described by a fit to
\begin{eqnarray}
\frac{m_{PS}}{2\pi T}&=&1+p_2 \,\hat{g}^2(T) + p_3 \, \hat{g}^3(T)+p_4 \,\hat{g}^4(T)\;,\nonumber \\
\frac{m_{V}}{2\pi T}&=&\frac{m_{PS}}{2\pi T} + s_4\, \hat{g}^4(T)\;,
\label{eq:psv}
\end{eqnarray}
where $\hat{g}^2(T)$ denotes the temperature-dependent running coupling renormalized in the $\overline{\mathrm{MS}}$-scheme 
at $\mu=2\pi T$. The perturbative value of $p_2$ from \cite{Laine:2003bd} is fully confirmed, while $p_3,p_4,s_4$ are not yet 
available analytically. 
Note that all coefficients are 
numbers, and the only temperature dependence of \eq(\ref{eq:psv}) resides in the coupling, whose logarithmically slow running is 
responsible for
the flat behavior observed for $T\gsi 1$~GeV in \fig\ref{fig:screen}. 
The spin dependence is found to be consistent with a single $O(\hat{g}^4)$ term $s_4$ over the entire temperature
range down to 1 GeV, and vanishes only for $T\rightarrow \infty$ with the running coupling.
Thus, (neglecting the wiggles within errors) all structure of the lattice data above 
$T\gsi 1$ GeV in \fig\ref{fig:screen} can be described by a sufficiently deep, resummed perturbative expansion 
about partonic degrees of freedom, and is
therefore characteristic of a quark gluon plasma.

What has remained entirely uncommented in the literature so far is the
rapid bending of the curves within $T\approx 0.5-0.7$~GeV, from a steep increase with temperature to an entirely flat behavior. 
The nearly vertical portions of the plot cannot possibly be accounted for by series like \eqs(\ref{eq:psv}), 
since their temperature dependence resides in the coupling only. 
The same bending is observed in the same
temperature range for \textit{all} $J=0,1$ mesons composed of $\bar{u}s$ and $\bar{s}s$ quarks as well  \cite{Bazavov:2019www}. 
That is, altogether this abruptly bending structure is present across 12 different quantum number channels! Since these constitute the 
dominant contributions to the partition function \eq(\ref{eq:ham}), 
an apparent change of dynamics takes place for the entire system, 
signalled by the complete breakdown
of resummed perturbation theory at the ``knee'' of those curves. At the temperatures in question, this cannot be caused
by chiral symmetry breaking. 
Rather, when decreasing temperature from the plasma regime, at the ``knee'' of the screening masses 
the chiral spin symmetric regime is entered,
which a perturbative calculation about partons cannot reproduce to \textit{any} order.  

Conversely, increasing temperature from the hadronic regime, each meson screening mass $m_\Gamma$ 
enters the perturbative regime at some individual screening temperature $T_\mathrm{s}(\Gamma)$, which one may define by, e.g., 
the most negative curvature of $m_\Gamma/T$ (the location of the bend),
\beq
 T_\mathrm{s}(\Gamma):\quad \min_{T}\left\{\frac{d^2}{dT^2}\frac{m_\Gamma}{T}\right\}\;.
 \eeq
Thus, for $T\gsi T_\mathrm{s}(\Gamma)$ quark hadron duality is realized in that quantum number channel\footnote{ The screening
masses discussed here were extracted by $\exp(-m_{scr} z)$ fits to the large distance correlators, which is appropriate for
bound states of $H_z$. For either unstable or multiparticle states, the exponential gets modified by
power law factors, whose general effect is a lowering of the resulting mass. While this implies some uncertainty
on the values of $T_\mathrm{s}(\Gamma)$, the exponential fits provide lower bounds on the true values.}. 
We may then conclude
that the bound states have released their quark gluon content, i.e.,
their chromoelectric interaction is screened. 
Once this happens in sufficiently many quantum number channels, chiral spin symmetry is broken 
as expected for a quark gluon plasma.  Note that the resulting value of $T_s$, where this happens, depends
on the precise flavor and mass content of the theory, as well as on the definition of $T_\mathrm{s}(\Gamma)$,
as expected for a crossover. 
 
We conclude that the behavior of meson screening masses from 12 different quantum number channels in $N_f=2+1$ QCD 
provide an independent demonstration of the existence of a temperature window $T_\mathrm{ch}\lsi T\lsi T_\mathrm{s}$,
in which chiral symmetry is restored but the dynamics is inconsistent with a partonic description.
By \eq(\ref{eq:ham}), it is then equally impossible to describe the equation of state in this regime by parton dynamics.

By contrast, \fig\ref{fig:screen} (left) shows chiral symmetry restoration to be achieved by
the initially heavier chiral partners of the lowest screening masses dropping abruptly around $T_\mathrm{ch}$, and the
same is true for all other flavour combinations  \cite{Bazavov:2019www}.
Then \eq(\ref{eq:ham}) implies growing pressure around $T_\mathrm{ch}$, also in the absence of parton dynamics. 
The same observation was made for chiral multiplets of baryons extracted from temporal lattice correlators. 
When used in a  hadron resonance gas calculation, these equally lead to growing pressure \cite{Aarts:2018glk} at and above
$T_\mathrm{ch}$.
  
\section{Chiral spin symmetry at finite temperature and density \label{sec:tmu}}

Having discussed the chiral spin symmetric temperature range $T_\mathrm{ch}\lsi T\lsi T_\mathrm{s}$ at zero density, 
what are the nature and values of its boundaries at $T_\mathrm{ch},T_\mathrm{s}$?
It is well-known that chiral symmetry restoration, marking $T_\mathrm{ch}$, 
proceeds gradually by an analytic crossover. On the other hand, very little attention has
been paid to the temperature range around $T_\mathrm{s}$ so far. The screening mass data of \fig\ref{fig:screen} 
suggest a smooth crossover as well. In order to rule out a non-analytic phase transition, a finite size scaling study would
be necessary to demonstrate that no discontinuity develops in the thermodynamic limit. 
It is well known that in the case of crossovers there are no
sharp phase boundaries, and thus the
numerical values for $T_\mathrm{ch},T_\mathrm{s}$ necessarily vary with their definition.
So far these are based on degeneracy patterns of meson correlators with fairly coarse temperature resolution, which is confirmed by 
the qualitative behavior of screening masses, as discussed in the last section. In addition,  $T_\mathrm{ch},T_\mathrm{s}$
depend on the number of quark flavors and their precise masses. 
For these reasons, accurate numbers for physical QCD are not available
at present. But they can be obtained straightforwardly, from standard lattice simulations of meson correlators, with 
high precision and temperature resolution at the physical point in the future.

The next question is what happens with this regime at non-vanishing
baryon chemical potential. The quark chemical potential term in the
QCD action is manifestly $SU(2)_{CS}$ and $SU(2N_F)$ symmetric \cite{Glozman:2017dfd}.
This suggests that both symmetries  observed at $\mu=0$ should also persist
at finite chemical potential.

It is well known from lattice simulations how the chiral crossover temperature, which constitutes
a lower bound for and is close to the chiral spin symmetric regime,
behaves for small $\mu_B\lsi 3T$. Several consistent evaluations give
\bea
\frac{T_\mathrm{pc}(\mu_B)}{T_\mathrm{pc}(0)}
&=&1-0.016(5)\left(\frac{\mu_B}{T_\mathrm{pc}(0)}\right)^2+\ldots\;,\nn\\
&\approx & \frac{T_\mathrm{ch}(\mu_B)}{T_\mathrm{ch}(0)}
\eea 
with the subleading term not yet statistically significant \cite{Bellwied:2015rza,Cea:2015cya,Bonati:2018wdn,Bonati:2018nut,HotQCD:2018pds}. 
The approximation in the second line is due to $T_\mathrm{ch}$ being somewhat larger than $T_\mathrm{ps}$, as discussed
in \Sec\ref{sec:obs} and visible in \fig\ref{fig:screen}. It can be improved upon by a suitable definition and quantitative evaluation
of $T_\mathrm{ch}$, e.g.~by $U(1)_A$ restoration. 

The qualitative behavior of the upper boundary can be inferred from
the value of a chosen meson screening mass at the temperature $T_\mathrm{s}$.
Here we choose vector mesons as they show the most pronounced bend across all flavor channels.
The screening mass at the bend corresponds to an inverse screening radius, 
\beq
r_V^{-1}\equiv m_V(\mu_B=0, T_\mathrm{s})=C_0T_\mathrm{s}\;.
\eeq
Beyond this length scale the corresponding screening mass behaves perturbatively,
i.e., for zero density the chromoelectric interaction is screened once $T>T_\mathrm{s}$.
Then, by $CP$-symmetry we know that mesonic screening
masses are even functions of $\mu_B/T$, and therefore
\beq
\frac{m_V(\mu_B)}{T}=C_0+C_2\left(\frac{\mu_B}{T}\right)^2+\ldots\;.
\label{eq:scr_mu}
\eeq
Keeping $r_V^{-1}$ constant as
chemical potential is varied, $dm_V\stackrel{!}{=}0$,
one finds
\beq
\frac{dT_\mathrm{s}}{d\mu_B}=-\frac{2C_2}{C_0}\frac{\mu_B}{T}-\frac{2C_2^2}{C_0^2}\left(\frac{\mu_B}{T}\right)^3+\ldots\;.
\eeq 
Since we know from analytic calculations \cite{Vepsalainen:2007ke} as well as 
lattice simulations \cite{Hart:2000ef,Pushkina:2004wa} that $C_2>0$, 
the upper boundary of the chiral spin symmetric regime leaves the temperature axis with zero slope and negative
curvature. We then conclude that the QCD phase diagram shows a chiral spin symmetric band that bends downwards with chemical
potential, as sketched in \fig\ref{fig:PD}. This can be checked straightforwardly 
by repeating the analysis of meson correlators 
from \cite{Rohrhofer:2019qwq,Rohrhofer:2019qal} with imaginary chemical potential. Since the chemical potential term is invariant under
chiral spin symmetry, screening masses
and correlators will be shifted differently for real and imaginary chemical potentials, cf.~\eq(\ref{eq:scr_mu}), 
but the degeneracy patterns should be the same in both cases.

\begin{figure}[t]
    \centering
    \includegraphics[width=0.98\columnwidth, clip, trim=3mm 2mm 0 0]{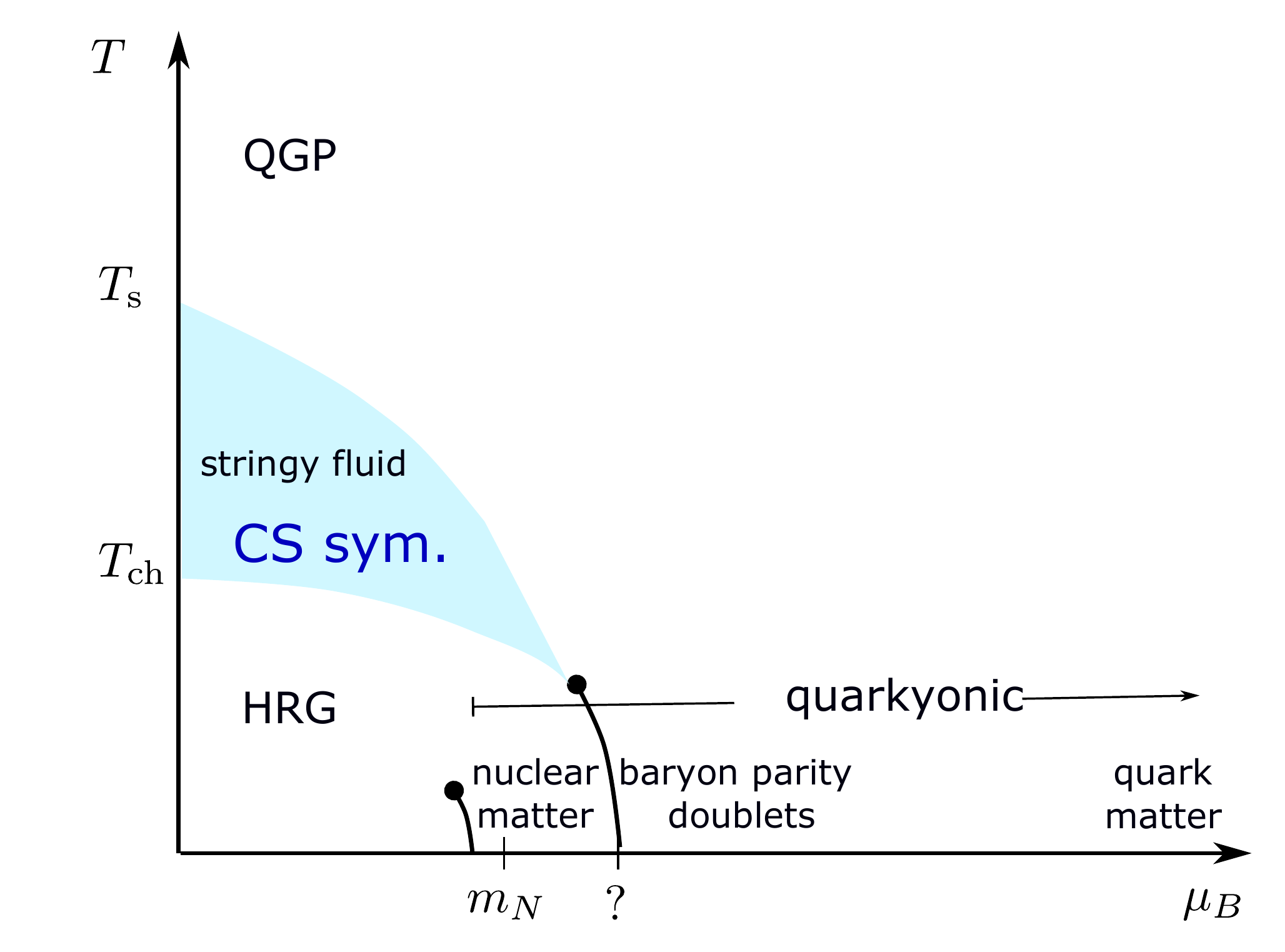}
    \caption{Qualitative sketch of a possible QCD phase diagram with a band of approximate chiral spin symmetry 
    termina\-ting at the critical
    end point of a non-analytic chiral phase transition.}
    \label{fig:PD2}
\end{figure}

Finally, we stress that these expectations concerning the shape of the chiral spin symmetric band hold for sufficiently
small $\mu_B/T$. The further qualitative behavior depends on the relative size of the curvatures $d^2T_\mathrm{ch}/d\mu_B^2$
and $d^2T_\mathrm{s}/d\mu_B^2$ with growing chemical potential. 
If the latter is sufficiently much larger than the former, then the boundaries
of the band merge at some non-vanishing temperature. This might in particular be expected to happen at the critical
endpoint of a possible first-order chiral phase transition, as sketched in \fig\ref{fig:PD2}. In this case the steeply rising  
part of the screening masses $m_\Gamma/T$, \fig\ref{fig:screen} right, would move towards $T_\mathrm{ch}(\mu_B)$ and evolve into a discontinuous jump as $\mu_B$ approaches
the critical endpoint, where the scalar screening masses have to vanish and display a kink. 
Again, at least a trend towards one or another behavior can be determined by screening mass
studies at imaginary chemical potential. 

\section{Baryonic parity doublet matter and its symmetries \label{sec:pdub}}

As chemical potential gets larger, we have no more reliable information from the lattice.
We now discuss on a merely qualitative level, 
how a chiral spin symmetric regime can exist in the baryon rich region at reasonably large
chemical potentials, as would be the case in a scenario like \fig\ref{fig:PD}.

It has been known for a long time that one can construct a manifestly
chirally symmetric Lagrangian with massive fermions if there
are degenerate fermions of opposite parity \cite{LEE}, the so
called parity doublet Lagrangian. Indeed, parity doubling of the light baryons is clearly
observed on the lattice above $T_\mathrm{ch}$  at zero density \cite{Aarts:2017rrl,Aarts:2018glk}, 
as a consequence of chiral symmetry restoration. 

Consider a pair of isodoublet fermion fields 
\begin{equation}
\Psi = \left(\begin{array}{c}
\Psi_+\\
\Psi_-
\end{array} \right),
\label{doub}
\end{equation}
where the independent Dirac spinors
$\Psi_+$ and $\Psi_-$ have positive and negative parity, respectively.
 Note that there
is in addition an isospin index which is suppressed.
The right- and left-handed fields are
directly connected with the opposite parity fields
\begin{equation}
\Psi_R = \frac{1}{\sqrt{2}}\left( \Psi_+ + \Psi_-\right);~~
 \Psi_L = \frac{1}{\sqrt{2}}\left( \Psi_+ - \Psi_-\right).
\label{RLL}
\end{equation}
The vectorial and axial parts of the chiral transformation 
 under the $(0,1/2) \oplus (1/2,0)$ 
 representation 
of $SU(2)_L \times SU(2)_R$ are
\begin{eqnarray}
\Psi &\rightarrow &
\exp \left( \imath \frac{\theta^a_V \tau^a}{2}
\otimes \mathds{1}\right)\Psi\;, \nonumber \\
\Psi &\rightarrow &
\exp \left(  \imath \frac{\theta^a_A\tau^a}{2}\otimes \sigma_1
\right)\Psi\;.
\label{VAD}
\end{eqnarray}
Here $\tau_j, \sigma_k$ are Pauli matrices that act in the  
spaces of isopsin and the parity doublets, respectively. In the  chiral transformation
law  the axial rotation  mixes two independent fields $\Psi_+$ and $\Psi_-$.
The chirally invariant Lagrangian of the free parity doublet is then given as
\begin{eqnarray}
\mathcal{L} &=& i \bar{\Psi}_+ \gamma^\mu \partial_\mu \Psi_+ + 
i \bar{\Psi}_- \gamma^\mu \partial_\mu \Psi_- \nonumber \\
&&- m \bar{\Psi}_+ \Psi_+ - m \bar{\Psi}_- \Psi_-\;.  
\label{lag}
\end{eqnarray}
The equivalence of the present
form of the Lagrangian \cite{LEE} with  those used in  \cite{Detar:1988kn,Jido:2001nt} was demonstrated in 
\cite{Catillo:2018cyv} and the chiral transformation law \eq(\ref{VAD}) corresponds
to the ``mirror" assignment of  \cite{Jido:2001nt}. 
In terms of the right- and left-handed fields of \eq(\ref{RLL}), we have
\begin{eqnarray}
\mathcal{L} 
 &=&  i \bar{\Psi}_L \gamma^\mu \partial_\mu \Psi_L + 
i \bar{\Psi}_R \gamma^\mu \partial_\mu \Psi_R \nonumber \\
&&- m \bar{\Psi}_L \Psi_L - m \bar{\Psi}_R \Psi_R\;.
\label{llag}
\end{eqnarray}
The latter form demonstrates  that the right-
and left-handed degrees of freedom are 
decoupled and the Lagrangian with massive fermions is  chirally invariant.
 
A crucial property of the  Lagrangian (\ref{lag},\ref{llag}) is that  the fermions
$\Psi_+$ and $\Psi_-$ are  degenerate and
have a nonzero chiral-invariant mass $m$.
The diagonal axial charge of the fermions $\Psi_-$ and $\Psi_+$ vanishes, while the off-diagonal
axial charge is 1. 
 
This Lagrangian can be supplemented by  
the pion and
sigma-fields of the linear sigma model \cite{Detar:1988kn,Jido:2001nt}.
Chiral symmetry breaking, $\langle 0|\sigma|0\rangle \neq 0$, generates a mass
splitting of the positive and negative parity baryons. I.e. the chiral symmetry
of the Lagrangian (\ref{lag}-\ref{llag}) is lifted. 
This regime
can be associated with nuclear matter, where physics is guided
by a coupling of nucleons of positive parity with $\pi,\sigma$
fields \cite{Floerchinger:2012xd,Drews:2016wpi,Brandes:2021pti}. 
The combined Lagrangian has sometimes been used 
in baryon spectroscopy \cite{Gallas:2009qp,Olbrich:2015gln} and as  a  model for chiral symmetry 
restoration scenario at high temperature or density,  where  baryons
with non-zero mass do not vanish upon a chiral restoration, 
see e.g.~\cite{Zschiesche:2006zj,Steinheimer:2011ea,Sasaki:2017glk,Larionov:2021ycq} 
and references therein.
Depending on the parameters the chiral restoration transition  
can be either of first or second oder \cite{Zschiesche:2006zj}. 

It turned out, however, that the free parity doublet Lagrangian 
 (\ref{lag}-\ref{llag}) has a larger symmetry than the 
$SU(2)_L \times SU(2)_R$ symmetry. It is in addition
manifestly $SU(2)_{CS}$ and $SU(2N_F)$ symmetric \cite{Catillo:2018cyv}.
Indeed, given  
\eq(\ref{RLL}), the parity doublet
(\ref{doub}) can be unitarily transformed into
a doublet
\begin{equation}
\tilde{\Psi} = \left(\begin{array}{c}
\Psi_R\\
\Psi_L
\end{array} \right)\;,
\label{doubtr}
\end{equation}
which is a two-component spinor composed of Dirac bi-spinors
$\Psi_R$ and $\Psi_L$ (i.e., altogether there are  eight components).

The Lagrangian (\ref{lag}-\ref{llag}) is obviously invariant under
the $SU(2)_{CS}$ rotations (\ref{eq:RL}),
\begin{equation}
\left(\begin{array}{c}
\Psi_R\\
\Psi_L
\end{array}\right)\; \rightarrow
\exp \left(i  \frac{\varepsilon^n \sigma^n}{2}\right) \left(\begin{array}{c}
\Psi_R\\
\Psi_L
\end{array}\right)\; .
\label{eq:su2cstra}
\end{equation}
Then it follows that the parity doublet Lagrangian is not only
chirally invariant under the transformation (\ref{VAD}), but also
$SU(2)_{CS}$- and $SU(2N_F)$-invariant with the generators of $SU(2N_F)$
being 
\begin{align}
\{
(\tau^a \otimes \mathds{1}),
(\mathds{1} \otimes \sigma^n),
(\tau^a \otimes \sigma^n)
\}.
\end{align}
We then conclude that baryonic parity doublet matter 
is a very natural candidate for a chiral spin symmetric regime at low temperatures.

\begin{figure}[t]
    \centering
    \includegraphics[width=0.5\columnwidth]{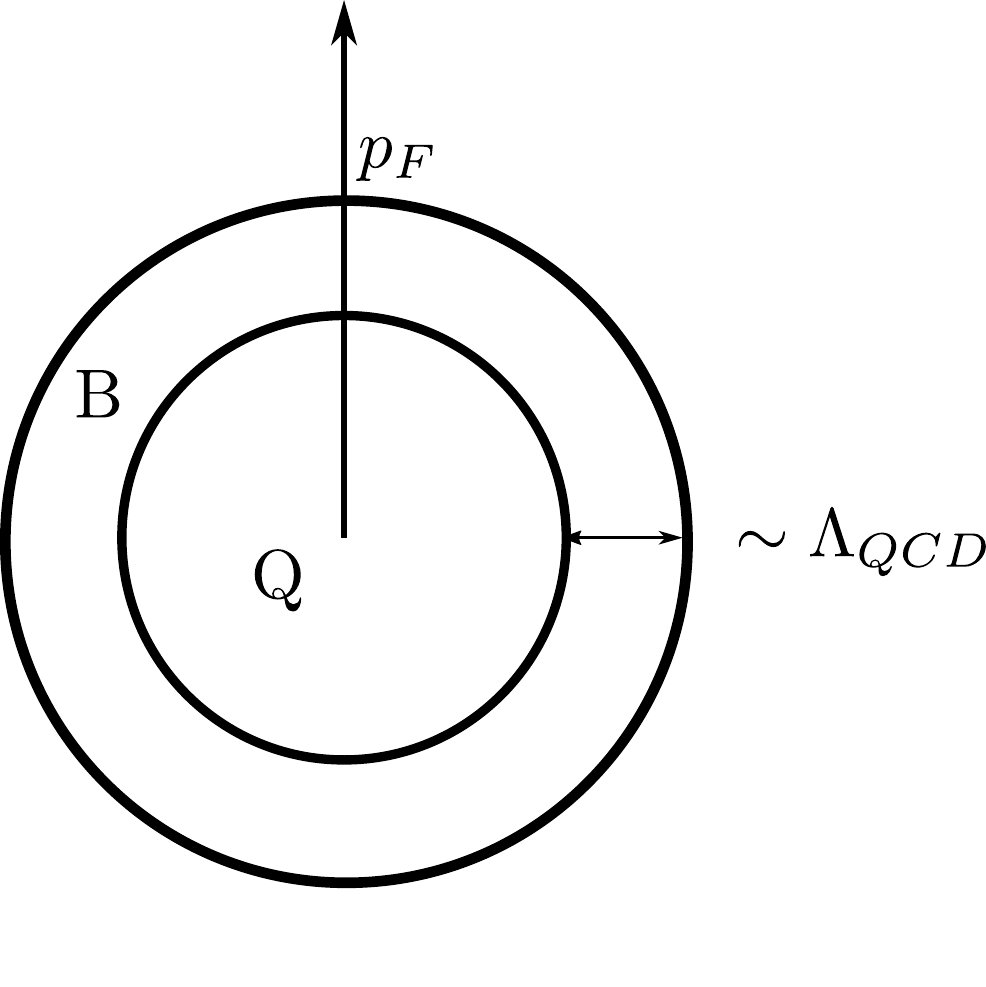}
    \caption{Qualitative sketch of the Fermi sphere of quarkyonic matter. The Fermi momentum is set by the chemical potential,
                  $p_F\sim\mu_B$. Quark matter inside is surrounded by a shell of baryon matter of 
                  thickness $\sim \Lambda_\mathrm{QCD}$, with a smooth, gradual transition between the two.}
    \label{fig:quarky}
\end{figure}
It is important to stress that
a coupling of pions and sigmas to the parity doublets
lifts the $SU(4)$ symmetry of the free parity doublet Lagrangian
because the $\pi-\sigma$-Lagrangian, while chirally invariant, is not
a $SU(4)$ singlet. If one insists on the SU(4) symmetry, then the usual 
pion-sigma Lagrangian must approximately decouple from the parity doubled
baryons. The $J=0$ mesons then reflect the usual chiral symmetry, but do not
mix with the proper $SU(4)$ multiplets to leading order, similar to the situation at $\mu_B=0$.
In summary, if parity doubled baryons decouple to leading order from $\pi,\sigma$, they are
chiral spin and $SU(4)$-symmetric. This corresponds to \fig\ref{fig:PD}.
Parity doublets strongly coupled to $\pi,\sigma$ are chiral
invariant but not chiral spin and $SU(4)$ symmetric., this corresponds to \fig\ref{fig:PD2}.

$SU(4)$-symmetric parity doubled baryon matter is fully consistent with the concept of quarkyonic matter, 
which was proposed in \cite{McLerran:2007qj} based
on large $N_c$ arguments. It characterizes the cold and dense regime of QCD and features the possibility 
of chirally symmetric but confined baryonic matter. Its defining features are the
pressure scaling as $p\sim N_c$ (for large $N_c$) and a Fermi sphere in momentum space, \fig\ref{fig:quarky}, consisting
of quark matter surrounded by a shell of baryons with thickness $\sim\Lambda_{QCD}$. Since the Fermi momentum is set
by the baryon chemical potential, $p_F\sim \mu_B$, quarkyonic matter interpolates between purely baryonic matter for 
$\mu_B\sim \Lambda_{QCD}$, and quark matter for $\mu_B\gg \Lambda_{QCD}$. The pressure scaling as well as the possibility 
of such a shell structure were verified for lattice QCD with heavy quarks by a combined strong coupling and 
hopping expansion \cite{Philipsen:2019qqm}, with the onset transition to baryon matter identified as the lower
boundary of this regime\footnote{For physical quark masses, the quarkyonic regime may be preceeded by a narrow regime 
of dilute baryon gas \cite{McLerran:2007qj}, but so far no QCD calculations are available on this.}.
Note also, that recent model investigations suggest neutron star data to be well described by 
baryonic matter up to six times nuclear density \cite{Brandes:2021pti}. In principle, this baryonic matter could then be
either chirally broken or symmetric.

Cold baryon matter can then appear in two different forms: ordinary nuclear matter with broken chiral symmetry, and
baryon parity doublet matter with restored chiral and approximate chiral spin symmetry, as sketched in \fig\ref{fig:PD}. 
These are separated
by the chiral phase transition or crossover. As the chemical potential increases, a growing fraction of the Fermi sea     
consists of quark matter and chiral spin symmetry is lost again. Beyond this density, the system is dominated by parton dynamics, 
i.e., quark matter. 
Currently, based on QCD nothing can be said about either the location or the width of the chiral spin symmetric band in 
the cold and dense regime, except that it can exist by continuation from the thermal regime.

\section{Dileptons and the chiral spin symmetric band \label{sec:dilep}} 

Dileptons have been used for a long time as a diagnostic
tool to study the nature of hadronic matter.
In vacuum the electron - positron annihilation into hadrons
shows a powerful resonance peak related to the existence
of $\rho$- and $\omega$-mesons, which reflects the
confining and chiral symmetry breaking properties of
the QCD vacuum. These properties persist also in a hadron resonance gas
or in a sufficiently dilute baryonic medium. Hence, experimental studies 
in heavy ion collisions at different
temperatures and chemical potentials
employ 
the inverse process, with the final state being
the electron-positron pair, to shed light on the question
to what extent  a hot or dense medium differs from the
vacuum. 

The differential dilepton production rate is determined by the spectral function of 
the electromagnetic current in the medium, expressed by its self-energy \cite{Kapusta:2006pm},
\begin{equation}
\frac{dN}{d^4q d^4x}= - \frac{\alpha^2}{\pi^3 M^2} f^{B}(q_0,T)
\mathrm{Im} [\Pi_\mathrm{em}(M,q;T,\mu_B)],
\end{equation} 
where $M$ is the invariant mass of the $e^+e^-$ pair with the four-momentum
$ q= (q_0, \vec q)$, $f^{B}(q_0,T)$ is the Bose-Einstein distribution
characterizing the thermalized me\-dium and $\alpha$ the fine structure constant.

The absence of resonance peaks is usually taken as a signal of chiral symmetry
restoration and deconfinement. For a detailed interpretation, however, some care is in order. 
For example, the perturbative
description of the electron-positron annihilation into hadrons
in vacuum  via $e^+ + e^- \rightarrow \bar q + q$
above the $ \rho, \omega, \phi$ resonance peaks 
receives so-called duality violating corrections. These are due to 
the very broad and overlapping $\rho', \rho'', ...$ resonances 
and/or instanton effects, which violate a purely partonic description and lead to the observed oscillating
behaviour of the corresponding spectral function \cite{Shifman:2000jv}.

Via \eq(\ref{eq:corr}), the finite temperature $\rho$- and $\omega$-meson
spectral functions are also encoded in the
Euclidean correlators of the vector currents
$\rho_{(1,0)+(0,1)}$ and $\omega_{(0,0)}$ of \fig\ref{fig:temp}.
Their extraction
from a Euclidean correlator with a finite number
of lattice points is an ill-defined problem. However, if a Euclidean correlator evaluated in full QCD is
very different from that calculated with non-interacting
quarks, one can safely state that the spectral
density  will not be dual to a perturbative description, but should
contain some remnant resonance structure. According to \fig\ref{fig:temp}, we expect this to be the case for the 
spectral functions of $J=0$ operators. 
On the other hand, the electromagnetic current correlator $\rho_{(1,0)+(0,1)}$
turns out to be close, but not equal, to the correlator calculated with free quarks. 
This could be caused by a fast decay of the $J=1$ excitation into $J=0$ excitations, $\rho\rightarrow \pi+\pi$. 
Hence, in this case the spectral density might not show an obvious resonance
structure, and the dilepton production should be close to the 
perturbative $ \bar q + q \rightarrow e^+ + e^-$ processes.
This is consistent with the less pronounced or absent $\rho$-peaks in the spectral function representing the fireball
above the chiral restoration temperature, as observed at RHIC \cite{STAR:2015zal,PHENIX:2015vek}, 
SPS \cite{NA60:2006ymb,CERES:2006wcq} and LHC \cite{ALICE:2018ael}.
We thus conclude that 
the absence of $\rho$ and $\omega$ peaks in high temperature dilepton spectra 
is entirely consistent with the chiral spin and $SU(4)$ symmetry above the chiral restoration.
Approximate $SU(4)$ symmetry requires the isoscalar $\omega_{(0,0)}$ correlator to be very
close to the isovector correlator $\rho_{(1,0)+(0,1)}$. Hence, what was said about
the absence of the $\rho$ peak above chiral
restoration line, should also be true with respect to the
$\omega$ peak. 

\begin{figure}[t]
    \centering
    \includegraphics[width=0.9\columnwidth, clip, trim=3mm 2mm 0 0]{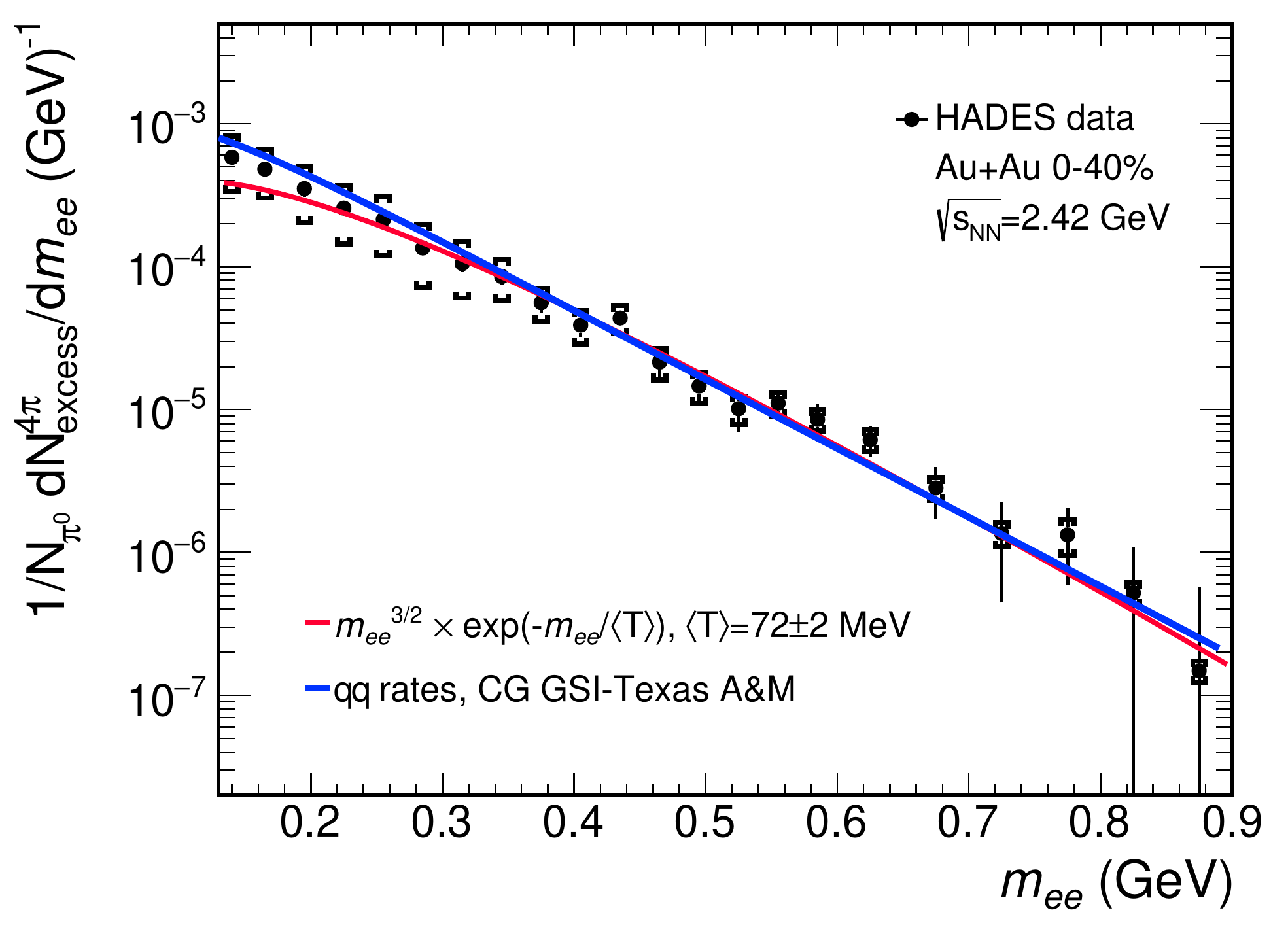}
    \caption{Acceptance-corrected dilepton excess yield obtained in Au-Au collisions at $\sqrt{s_{NN}}=2.42$ GeV \cite{HADES:2019auv}.}
    \label{fig:hades}
\end{figure}

A remarkable piece of information about the phase structure at large chemical
potentials and low temperatures is delivered by recent experimental results from the HADES collaboration \cite{HADES:2019auv}.
HADES measured the dielectron production in Au-Au collisions
at $\sqrt{s_{NN}}=2.42$~GeV. The excess yield extracted by subtracting
the $\eta, \omega$ contributions, which are produced beyond the fireball, 
is shown\footnote{We thank Tetyana Galatyuk and the HADES collaboration for preparing \fig\ref{fig:hades}}
in \fig\ref{fig:hades}. It exhibits a nearly exponential fall-off.
A fit of the black-body spectral distribution
\beq
\frac{dN}{dM} \sim M^{3/2} e^{-M/T}
\eeq 
to the data (red curve) yields $T \sim 72$ MeV.  No pronounced $\rho$-structure is visible,
and the data are well described by the leading order $ \bar q + q \rightarrow e^+ + e^-$
diagram (blue curve).
A slight oscillation
about the perturbative curve seems apparent in \fig\ref{fig:hades}, hinting at hadronic duality violations.
Such violations can appear both in the chirally broken and in the chirally symmetric phase.

The HADES result is also consistent with the in-medium broadening of $\rho$-mesons
(for the corresponding curves, see ref. \cite{HADES:2019auv}). The Rapp-Wambach model \cite{Rapp:1999ej}
describes coupling of pions with the nucleon - nucleon-hole excitations in the baryon medium
(including also different baryon resonances). The $\rho \rightarrow \pi + \pi$ width then
becomes very broad because of the additional couplings of the $\rho$-meson to these collective excitations.
This mechanism implicitly relies on spontaneous breaking of chiral symmetry,
since its parameters are extracted from the 
elementary processes in vacuum. 

A third possibility for the disappearance of the $\rho$-peak
in a dense chirally symmetric baryonic medium 
is provided by chiral spin and $SU(4)$ 
symmetric parity doublet matter. 
When chiral symmetry is restored, the baryons of positive and
negative parity to leading order decouple from pions and sigmas. Hence a baryonic medium becomes
a true Fermi gas.
An electromagnetic coupling of  baryon - baryon hole transitions to
photons guarantees
 an equilibrium between the baryonic Fermi gas and the photonic  Bose
gas. Photons can be converted into electron-positron pairs.
This is consistent with the black-body radiation description
of the excess shown in \fig\ref{fig:hades}. The dilepton production
of baryonic parity doublet matter is hence very similar to that of  thermalized
quark matter.
This suggests that the HADES point at $T \sim 72$ MeV, $\mu_B \sim 900$ MeV
is just above the
chiral restoration line, $T_\mathrm{ch}(\mu_B)$, and could possibly be within the
chiral spin and $SU(4)$ symmetric band. It would be most interesting to 
further assess whether there is a fine structure in the dilepton data allowing to further
distinguish between scenarios such as those sketched in \figs\ref{fig:PD} and \ref{fig:PD2}. 

 \section{Conclusions}
 
Based on recent lattice studies of meson correlators with chiral fermions, we must take 
approximate $SU(2)_{CS}$  and $SU(4)$ symmetries of QCD, as well as its associated dominance of
color-electric interactions between light quarks, as a matter of fact  in a temperature range above the chiral
crossover. In this work we have collected already published spectra of screening masses, 
and employed their direct relation to the QCD partition function to confirm the existence of such an 
intermediate temperature regime independent of symmetry arguments. 

In particular, there is non-perturbative evidence that color-electric fields are dynamically
enhanced over color-magnetic ones for the temperatures in question. Debye screening is a non-relativistic
concept which applies to heavy quarks, but its associated Debye mass does not enter the partition function
for $N_f=2+1$ QCD with relativistic quarks. We then applied quark hadron duality as a 
criterion for screening: once a hadron screening mass in a given quantum number channel becomes amenable
to a perturbative description, the color-electric interaction between quarks is sufficiently screened to allow for parton dynamics.
Recent lattice results show this to happen in a narrow temperature window $T\sim 0.45-0.75$ GeV across
12 different quantum number channels of $J=0,1$ mesons, with partonic behavior for higher and non-perturbative
behavior for lower temperatures. 

From the known behavior of meson screening masses with small baryon chemical potential, we then conjectured
the chiral spin symmetric band to extend downwards across the QCD phase diagram.
At least the behavior for small baryon densities can be fully determined by future studies of spatial correlation functions
at imaginary chemical potential with known techniques.
Baryonic parity doublet matter satisfies both chiral spin and $SU(4)$ symmetries, and is a candidate
for the form of nuclear matter beyond the chiral crossover/phase transition at $T=0$. 
In the cold and dense regime however, the location and width of
a chiral spin symmetric regime is not yet constrained from QCD. 
High precision dilepton spectra with fine resolution, combined with theoreti\-cal efforts towards QCD spectral functions, 
should be a promising tool to unravel the nature of the degrees of freedom in the chiral spin symmetric regime.\\

\noindent
{\bf Note added:} After completion of our 
manuscript an analysis of the pseudo-scalar spectral function at temperatures
above the chiral crossover appeared \cite{Lowdon:2022xcl}. It
shows resonance-like peaks for the pion and its first excitation
which only gradually disappear, consistent with the non-perturbative dynamics of the 
chiral-spin-symmetric band, as desribed here.\\

\section*{Acknowledgments:}
        We thank Tetyana Galatyuk, Ralf Rapp, Giorgio Torrieri and Wolfram Weise for discussions.
        O.~P.~acknowledges support by the Deutsche Forschungsgemeinschaft (DFG) 
        through the grant CRC-TR 211 ``Strong-interaction matter
    under extreme conditions'', as well as by the State of Hesse within the Research Cluster ELEMENTS (Project ID 500/10.006).

\bibliographystyle{utphys}
\bibliography{./references}

\end{document}